\documentclass{article}

\def\be{\begin{equation}}
\def\ee{\end{equation}}
\def\ba{\begin{array}}
\def\ea{\end{array}}
\def\de{\partial}

\def\pmb#1{\setbox0=\hbox{$#1$}%
  \kern-.025em\copy0\kern-\wd0
  \kern.05em\copy0\kern-\wd0
  \kern-.025em\raise.0433em\box0}

\begin{document}

\title{5-dimensional special relativistic hydrodynamics and cosmology}

\author{Gianluca Gemelli}
\date{{\small L.S. B. Pascal, V. P. Nenni 48, Pomezia (Roma), Italy. \\
e-mail: gianluca.gemelli@poste.it}}

\maketitle

\begin{abstract}
5-dimensional special relativity can be considered as the 5-dimensional extension of Carmeli's cosmological special relativity, as well as the flat specialization of 5-d brane world theory. 
To this framework we add a 5-dimensional perfect fluid stress-energy tensor, and unify the equations of perfect hydrodynamics in a single 5-dimensional tensor conservation law. This picture permits to interpret particle production phenomena as cosmological effects, in the spirit of open system cosmology.
The source of particle production vanishes if the fluid is
isentropic. Moreover the hydrodynamical equations can be interpreted in terms of a scale factor, giving
rise to a set of equations which simulate in a sense Friedmann cosmology. 
\end{abstract}

\section{Introduction}

To what extent the dynamics of the mass-energy content of the Universe can be described in special relativistic terms? 

In this paper we deal with the 5-dimensional generalization of the equations of special relativistic hydrodynamics which were studied recently in \cite{gemelli,gemellibis} and its consequences of cosmological interest: a particle production phenomenon and a set of evolution equations for the universe expansion. 

The most natural interpretation of that picture is in the framework of
cosmological special-relativity, in conformity with Carmeli's theories 
\cite{carmeli95,carmeli96,carmelibook}, with the receding velocity of galaxies playing the role of fifth dimension. 

However, as for the particle production mechanism described here, there is no need to suppose the fifth dimension to be timelike, like in cosmological relativity. The signature of the fifth dimension is instead involved in the Friedmann-like universe evolutions equations, and selects in a sense the type of the universe.

Modern relativistic inflationary cosmology usually involves particle production (see e.g. \cite{cissoko98}). Cosmological particle production is thought in fact to account for negative pressure, which arises in the modelling of the accelerating universe \cite{decampos}. The cosmological scenario with particle production is often called open system cosmology \cite{prigogine}.

Here we interpret particle production phenomena as a genuine cosmological effect, contained in a model which links
cosmological relativity (or rather, the 5-d version of it) and open system cosmology. This is interesting, since in cosmology particle production is usually introduced by hand; the 5-d hydrodynamic approach instead has it as a natural consequence of 5-d cosmology, and, rather unespectedly, even has it in the limit of negligible gravitation. 

The paper is organized as follows.

In section 2 we give a brief review of relativistic hydrodynamics, which will be useful later on. 

In section 3 we introduce the general framework of 5-d special relativity with 5-d stress-energy conservation.

In section 4 we specialize the 5-d stress-energy tensor to the case of perfect hydrodynamics.

In section 5 we study the particle production phenomenon contained in the model above. We find that the source of particle production vanishes if the fluid is isentropic.

In section 6 we see that the 5-d cosmological special relativistic hydrodynamical equations lead to dynamical equations in terms of a scale factor, equations which resemble or are analogues in a sense to those of Friedmann's.

\section{Relativistic hydrodynamics}

Let ${\cal M}_4$ denote the minkowski space-time, of signature $- + + +$. 
Let greek indices run from $0$ to $3$. Units are chosen in order to 
have the speed of light in empty space $c\equiv 1$.

A relativistic continuum system is characterized by a world tube 
$\Omega\subset {\cal M}_4$, generated by the set of the 
worldlines of its constituting elementary particles, 
with tangent unit timelike vector $U^\alpha$.
The world tube $\Omega$ is the support of a stress-energy
tensor $T^{\alpha \beta}$, which satisfy the conservation
law: $\de_\alpha T^{\alpha\beta}=0$. 

For a perfect fluid the matter-energy tensor is a function of
the dynamical and thermodynamical variables (see for ex. 
\cite{lich67libro,fer85,anile89,lich94}):

\be
T^{\alpha\beta} = (\rho +p)U^\alpha U^\beta + p \eta^{\alpha\beta}
\label{T}
\ee
where $\eta^{\alpha\beta}$ is the Minkowski metric, $\rho \ge 0$ is the proper energy 
density and $p \ge 0$ the proper pressure. We suppose $\rho + p>0$. 
We moreover set 
\be
\rho = r(1+{\cal E})
\label{rE}
\ee 
where $r\ge 0$ is the matter density (baryon number) and ${\cal E}\ge 0$ 
the internal energy. Let us introduce the proper 
temperature $T$,
the specific entropy $S$ and the thermodynamic principle
\be
T dS = d{\cal E}  -\frac{p}{r^2} dr
\label{tp}
\ee
It is convenient to adopt $p$ and $S$ as
the fundamental thermodynamical variables, and to introduce a generic
equation of state of the form: $r=r(p,S)$. Thus from (\ref{tp}) we  
have  two independent relations: $T=(\de {\cal E}/\de S) -pr^{-2}(\de r/dS)$ 
and $T (\de S/\de p)= (\de {\cal E}/\de p) -pr^{-2}(\de r/dp)$. 
We therefore
have 5 independent variables: $p$, $S$ (or any other pair of 
thermodynamical variables) and the three independent components of 
$U^\alpha$ ($U^\alpha U_\alpha =-1$).

The differential system of relativistic hydrodynamics is the following:

\be
\ba{ll}
\de_\alpha (r U^\alpha)=0, & \de_\alpha T^{\alpha\beta}=0
\ea
\label{system}
\ee
Here (\ref{system}$)_1$ means conservation of the matter density (which 
corresponds to the conservation of the specific number of particles); 
therefore nuclear reactions of particle creation are neglected within this scheme. As for
(\ref{system}$)_2$, it 
is the ordinary conservation equation for the energy
tensor of the fluid. The total number of scalar equations of 
(\ref{system}) is 5.

An equivalent formulation of (\ref{system}) is obtained by
replacing equation
$\de_\alpha (r U^\alpha)=0$ with $U^\alpha \de_\alpha S=0$. 
However formulation (\ref{system}) is preferable
for its conservative form.

\section{5-d special relativity}

Let us briefly recall the notations and hypothesis introduced in \cite{gemelli,gemellibis}, with some important completion.

Let us consider a 5-dimensional flat manifold ${\cal M}_5$. We leave for the moment the possibility for the signature to be $- + + + -$ or $- + + + +$ as well; to this aim we will introduce in the equations a scalar $\epsilon$ which can assume the values $+1$ or $-1$. Let capital latin indices 
run from $0$ to $4$ and greek letters run from $0$ to $3$. 
We choose orthonormal coordinates, such that the 5-dimensional 
line element is:

\be
g_{AB}dx^Adx^B
= -dt^2 + dx^2 +dy^2 +dz^2 +\epsilon d\xi^2
\ee
where we have denoted $t=x^0, x=x^1, y=x^2, z=x^3$ and $\xi=x^4$. 

In the case of cosmological special relativity we have $\epsilon=-1$ and $\xi=v$, where $v$ is the receding velocity of galaxies. In fact if $\epsilon=-1$ we may equivalently being considering the 5-d extension of cosmological special relativity or the flat-spacetime specialization of Carmeli's general relativistic brane world theory \cite{carmelibook}.   
In practice we set both the speed of light in vacuo $c=1$ and the Hubble-Carmeli constant (analogue to $c$ for the fifth dimension) $1/H_0=1$.

Let us now define 5-d perfect hydrodynamics in an axiomatic way, by analogy with 4-d hydrodynamics, and then study
the 4-d consequences of such choice. 
Let ${\cal T}^{AB}$ be a 5-d stress-energy tensor.

Then, let us introduce the constant unit vector 
$\Xi = \de_{x^4}$ of the fifth dimension;
we have $\Xi^A=\delta_4{}^A$, $\Xi_B=\epsilon \delta_B{}^4$ and 
$\Xi^A\Xi_A=\epsilon$. 

The (unique) splitting of 
${\cal T}^{AB}$ along the  $\xi$-direction and the orthogonal
complement of ${\cal M}_5$ (which is ${\cal M}_4$) is

\be
{\cal T}^{AB}=T^{AB} + {P^A}\Xi^B + P^B\Xi^A + E\Xi^A\Xi^B
\label{decoT}
\ee
where $T^{AB}$ and $P^A$ are orthogonal to $\Xi$ and $T^{AB}$ is
symmetric (it is the ordinary matter-energy tensor).

Let ${\cal T}^{AB}$ be conserved in ${\cal M}_5$, i.e. let us postulate
the following 5-dimensional conservation equation:

\be
\de_A {\cal T}^{AB}=0
\label{conserva5}
\ee
This actually is a 5-dimensional generalization of special relativistic
continuum dynamics.
The splitting of (\ref{conserva5}) along $\Xi$ and its orthogonal
complement gives rise to the following equivalent system

\be
\ba{ll}
\de_A T^{AB} + \Xi^A\de_A P^B =0, & \de_AP^A + \Xi^A\de_A E=0  
\ea
\ee
or, in our coordinates:
\be
\ba{ll}
\de_\alpha T^{\alpha\beta} + \de_\xi P^\beta =0, & 
\de_\alpha P^\alpha + \de_\xi E=0  
\ea
\label{5dsystem}
\ee

\section{5-d special relativistic hydrodynamics}

Let us now define, by analogy with the case of perfect hydrodynamics:

\be
{\cal T}^{AB}=(M+Q)V^AV^B + Q g^{AB}
\label{RQ}
\ee 
Here the ``thermodynamical" fields $M$ and $Q$ are supposed to be defined and regular in a 5-d ``world tube" generated by
a geometrical congruence of lines tangent to $V$. Let $s$ be a privileged parameter along one of such lines, which we denote by $\ell$, with parametric equations $x^A=X^A(s)$; we thus have, along $\ell$: $M=M(s)$, $Q=Q(s)$, and:

\be
V^A= \frac{dx^A}{ds}.
\ee
Note that in \cite{gemelli} the symbol $R$ was used in place of $M$; here we instead deserve $R$ for a more appropriate use (see section 4).
Now let us introduce the following splitting:

\be
V^A=W^A+\mu \Xi^A
\ee
where $\Xi^A=\delta^A{}_\xi$ is the direction of the fifth dimension and $\mu$ a free parameter, the square of which is equivalent to the square of $V$, as we are going to see. 
We have: $V^\alpha=W^\alpha=dx^\alpha/ds$ and $V^\xi=\mu=d\xi/ds$. 

Let us moreover denote by a star the derivative with respect to $s$ and by prime the derivative with respect to $\xi$, so that we have:

\be
(\ \ )^\star=V^A\de_A=W^\alpha \de_\alpha + \mu (\ \ )'.
\label{dds}
\ee
We also have: $V^A=(X^A)^\star$, $\xi^\star=\mu$.
Let us denote $E={\cal T}^{\xi\xi}$. Now system (\ref{conserva5}) equivalently reads:

\be
\ba{ll}
\de_\alpha {\cal T}^{\alpha\xi}+E'=0\cr
\de_\alpha T^{\alpha\beta}+({\cal T}^{\xi\beta})'=0
\ea
\label{5dhydro}
\ee
where:
\be
E=\frac{r^2}{\rho + p}+ \epsilon p.
\label{gemelli}
\ee
Recall that for an ordinary perfect fluid the 4-d system of hydrodynamics is (\ref{system}), i.e.:

\be
\ba{ll}
\de_\alpha (rU^\alpha)=0\cr
\de_\alpha [(\rho+p)U^\alpha U^\beta +p g^{\alpha\beta}]=0
\ea
\label{4dhydro}
\ee 
System (\ref{4dhydro}) must be considered united to an equation of state of the kind $p=p(r,S)$ 
and to the thermodynamical principle (\ref{tp}) holds.

To have a significant match between (\ref{5dhydro}) and (\ref{4dhydro}) we have to suppose:

\be
\ba{l}
\mu(M+Q)W^\alpha = rU^\alpha \cr
(M+Q)W^\alpha W^\beta +Qg^{\alpha\beta} = (\rho+p)U^\alpha U^\beta + p g^{\alpha\beta}
\ea
\ee
One thus necessarily finds:

\be
\ba{ll}
Q=p, & M=\frac{r^2}{\mu^2 (\rho+p)}-p
\ea
\label{comparison}
\ee
and consequently:

\be
W^\alpha = \mu \frac{\rho+p}{r}U^\alpha.
\ee
In relativistic hydrodynamics the variable 
$f=(\rho+p)/r$ 
is called fluid index, and it is $f=i+1$, where $i$ is the specific enthalpy (see \cite{lich94} p. 99).
We then have: 

\be
W^\alpha = \mu fU^\alpha.
\ee
We also have:

\be
V^2=V^AV_A=\mu^2\Bigl[\epsilon-f^2\Bigr]
\label{mumu}
\ee
By introducing the symbol $V^2$ we implicitedly assume $V_AV^A>0$. Even if we could do without such hypothesis, for the moment we prefer to work with $V_AV^A>0$ for the sake of simplicity. 
We see from (\ref{mumu}) that the square of the parameter $\mu$ is equivalent to the square of $V$, as said before. We leave $\mu$ as a free parameter for the moment. For the sake of brevity we denote by a dot derivative with respect to proper time, i.e.: $(\ \ )^{\pmb{\cdot}}= U^\alpha \de_\alpha$; this should not be confused with derivative with respect to coordinate time, as it is instead in (\ref{Fried1})-(\ref{Fried2}). 
From (\ref{dds}) we then have:

\be
(\ \ )^{\star}=\mu [ f(\ \ )^{\pmb{\cdot}} + (\ \ )^\prime ]
\label{ddsagg}
\ee
Now let us apply (\ref{ddsagg}) to $\xi$ and compare with $\xi^\star=\mu$; we are led to the following identity:
\be
f\dot\xi =0
\label{fifi}
\ee
We discard for the moment the singular situation $f=0$ (otherwise $\rho+p=0$) and conclude from (\ref{fifi}) that $\dot\xi=0$. 
Our hydrodynamical system now reads as follows:

\be
\ba{l}
\de_\alpha (rU^\alpha)+E'=0\cr
\de_\alpha T^{\alpha\beta}+(rU^\beta)'=0
\ea
\label{sist1}
\ee
In particular, the function $-E'$ is interpretable as the source of particle production. 

\section{Cosmological particle production}

Let us now introduce in (\ref{sist1}) the expression of the 4-d component of the stress-energy tensor:
\be
T^{\alpha\beta}=(\rho+p)U^\alpha U^\beta +p g^{\alpha\beta}
\ee
and split the system with respect to $U$ and the orthogonal local rest space \cite{manyfaces}; we have:

\be
\ba{l}
\dot r +r \de_\alpha U^\alpha + E' =0 \cr
\dot \rho + (\rho + p) \de_\alpha U^\alpha + r'=0 \cr
(\rho + p)\dot U {}^\beta + r (U^\beta)' + \de^\beta p + U^\beta \dot p =0
\ea
\label{sist2}
\ee
Note that if we remove from (\ref{sist2}) all the terms with a prime, i.e. we consider in a sense the ordinary 4-d situation, 
in wich all fields are independent on $\xi$, we obtain nothing but the ordinary 4-d hydrodynamical system (\ref{4dhydro}), i.e., with the dot notation:
\be
\ba{l}
\dot r +r \de_\alpha U^\alpha =0 \cr
\dot \rho + (\rho + p) \de_\alpha U^\alpha=0\cr
(\rho + p)\dot U {}^\beta  + \de^\beta p + U^\beta \dot p =0
\ea
\label{sist0}
\ee
From (\ref{sist2}$)_1$ and (\ref{sist2}$)_2$ we have:

\be
\dot\rho -f(\dot r + E')+r'=0
\label{Erela}
\ee
Now since we have by definition \cite{gemelli}:

\be
\rho=r(1+{\cal E})
\ee
where ${\cal E}$ is the internal energy, from the thermodynamical principle (\ref{tp}) we have:
\be
d \rho = fd r+ rT d S
\label{drho}
\ee  
so that from (\ref{Erela}) we have the following relation for $E'$:

\be
E'=f^{-1}(rT\dot S +r')
\ee
However, we have from (\ref{gemelli}): 

\be
E=\frac{r^2}{\rho+p}+\epsilon p=rf^{-1}+\epsilon p
\label{Edef}
\ee
and thus, from comparison of (\ref{Erela}) and (\ref{Edef}) we have:

\be
rf^{-1} T(\dot S+f^{-1}S')=(\epsilon-f^{-2})p'
\ee
or equivalently, from (\ref{dds}):

\be
r^2f^{-2}\mu^{-1}TS^\star=(\epsilon-f^{-2})p'
\label{dSs}
\ee
From (\ref{dSs}) we have that if all fields are independent on $\xi$, like in ordinary 4-d hydrodynamics, we have
$\dot S=0$, which in fact is a well known consequence of system (\ref{sist0}).

From (\ref{dSs}) it is also possible to conclude that particle production is absent if the fluid is isentropic, i.e. if $dS=0$ [and the equation of state consequently reduces to $p=p(r)$] then $E'=0$. In fact if $dS=0$ we have:

\be
(\epsilon-f^{-2})p'=0
\label{dSs0}
\ee
and therefore there are two possible situations: $p'=0$ or $\epsilon=f^{-2}$. 

If $p'=0$ then from the equation of state we also have $r'=\rho'=0$ and consequently $E'=0$. 

If instead $\epsilon=f^{-2}$, then we must have $\epsilon=+1$ and $r^2=(\rho+p)^2$. We consequently have:

\be
rdr=(\rho+p)(d\rho+dp)
\label{rrr}
\ee
Now from (\ref{tp}) if $dS=0$ we have $d\rho=r^{-1}(\rho+p)dr$ so that from (\ref{rrr}) we have:

\be
(\rho+p)dp=0
\ee
Excluding the singular case $\rho+p=0$ we conclude $dp=0$ and consequently $dr=d\rho=0$, which implies $dE=0$ and thus again $E'=0$.  

Thus in any case the source of particle production vanishes if $dS=0$.

\section{Friedmann cosmology}

Friedmann equations are (\cite{carmelibook} p. 168-169 and \cite{wesson} p. 15):

\be
\Biggl(\frac{\dot R}{R}\Biggr)^2=\frac{1}{3}(\chi\rho_F +\Lambda)-\frac{k}{R^2}
\label{Fried1}
\ee

\be
\frac{\ddot R}{R}=-\frac{\chi}{6}(\rho_F+3p_F) +\frac{\Lambda}{3}
\label{Fried2}
\ee
where $R$ is the scale factor, $\chi$ is the gravitational constant, $\Lambda$ is the cosmological constant, $k=+1,0,-1$ and $\rho_F$ and $p_F$ are density and pressure of the cosmological matter-distribution of curved spacetime. We had to introduce the suffix $|_F$ to distinguish them from the analogous fields of our test distribution in flat spacetime. In (\ref{Fried1})-(\ref{Fried2}) dots mean derivatives with respect to the coordinate time. 
 
The fact that equations (\ref{Fried1})-(\ref{Fried2}) can be formally obtained in a 5-d special relativistic framework, as we are going to see, is not a
completely surprising fact, since the full 5-d hydrodynamical system gives us free parameters to play with, but
it is also non trivial, since Friedmann equations are Einstein's gravitational equations, while we are working in a flat-spacetime, with no gravitation. 

The sgnificative idea then is that 5-d special relativistic hydrodynamics can simulate, at least to some extent, general relativistic 
cosmology.

Let us consider the general case $dS\not \equiv 0$, i.e. with a possibly nonvanishing source of particle production, and let us turn to the original 5-d system. In relativistic hydrodynamics the variable ${\cal T}={f}/{r}$
is called dynamical volume (see \cite{lich94} p. 99). Let  $\Phi=(\mu^2 {\cal T})^{-1}$. From (\ref{comparison}) we then have 
\be
M+Q=\Phi
\ee
It is a useful idea, on physical terms, to imagine that the dynamical volume should be proportional to the cube of a parameter $R$, representing a ``typical length", and that therefore our variable $\Phi$ should be proportional to $R^{-3}$; we will introduce this hypothesis later on. 

From (\ref{conserva5}) and (\ref{RQ}) we then obtain the following general form of the 5-d system:

\be
\Phi^\star V^B+\Phi (V^B)^\star +\Phi \de_AV^AV^B+\de^Bp=0
\label{systemgeneral}
\ee
We now are going to consider some useful consequences of system (\ref{systemgeneral}).
 
By multiplying (\ref{systemgeneral}) by $X_B$ we have:

\be
\Phi X_B(V^B)^\star +(\Phi^\star +\Phi \de_AV^A)X_BV^B +X^B\de_B p=0
\label{X}
\ee
By multiplying (\ref{systemgeneral}) by $V_B$ we have:

\be
\Phi V_B(V^B)^\star +(\Phi^\star +\Phi \de_AV^A)V_B V^B + p^\star =0
\label{V}
\ee
Finally, by taking the 5-d divergence of (\ref{systemgeneral}), i.e. in practice by multiplying it by $\de_B$, we have:
\be
\Phi^\star{}^\star+2\Phi^\star\de_AV^A +(V^B)^\star\de_B\Phi +\Phi [\de_B(V^B)^\star+(\de_BV^B)^\star]+\Phi (\de_AV^A)^2 + \de_A\de^A p=0
\label{de}
\ee
Now, since $V^B=(X^B)^\star$, we have:

\be
\ba{l}
X_BV^B=(X^BX_B)^\star/2 \cr
X_B(V^B)^\star = (X^BX_B)^\star{}^\star/2-V_BV^B \cr
V_B(V^B)^\star= (V_BV^B)^\star/2
\ea
\ee 
Moreover, since $\Phi=\Phi(s)$, we write: $(V^B)^\star \de_B \Phi=\Phi^\star \de_AV^A$. In fact:
$$ 
\Phi^\star (V^B)^\star  \frac{ds}{dX^B}=\Phi^\star \frac{(V^B)^\star}{(X^B)^\star}
$$
and we also have:
$$
\frac{(V^B)^\star}{(X^B)^\star} =\frac{dV^B}{ds}\frac{ds}{dX^B}=\frac{dV^B}{dX^B}
$$
Therefore, denoting, for the sake of brevity: $X^2=X_BX^B$, $\de_X=X^B\de_B$, $\Delta=\de_A\de^A$ and $ \nabla V=\de_AV^A$ we have that (\ref{X})-(\ref{V})-(\ref{de}) take the following form:

\be
(1/2)\Phi[(X^2)^\star{}^\star-2V^2]+(1/2)(\Phi^\star+\Phi \nabla V)(X^2)^\star +\de_Xp=0
\label{X1}
\ee

\be
\Phi^\star V^2+(1/2)\Phi(V^2)^\star +\Phi \nabla V V^2 + p^\star =0
\label{V1}
\ee

\be
\Phi^\star{}^\star+3\Phi^\star \nabla V + \Phi [\de_A(V^A)^\star + \nabla V^\star]+\Phi \nabla V^2 + \Delta p=0
\label{de1}
\ee
Note that, with our use of the symbol $X^2$ we again implicitedly assume, for the sake of simplicity, $X_AX^A>0$, which is restrictive, since $X^AX_A=x_ix^i -t^2-\xi^2$ in general could be non positive. However we may be considering $t=0$ (present time) and ``large distances" in a sense.    
 
Now let us introduce the typical length parameter $R$, in a crude and simple way, i.e. by taking  $X^2=R^2$ and $\Phi=R^{-3}$. This $R$ should not be confused with the Ricci scalar of general relativity. It also should not be taken as implying the existence of a physical boundary. From (\ref{X1})-(\ref{V1})-(\ref{de1}) we have respectively:

\be
\Phi[(R^\star)^2+RR^\star{}^\star-V^2]+(-3\Phi R^\star/R +\Phi \nabla V)RR^\star +\de_Xp=0
\label{X2}
\ee

\be
-3\Phi V^2R^\star/R+(1/2)\Phi(V^2)^\star +\Phi \nabla V V^2 + p^\star =0
\label{V2}
\ee

\be
-\Phi[3R^\star/R-12(R^\star/R)^2+9 \nabla V R^\star/R]+ \Phi [(\nabla V)^\star + \nabla (V^\star)]+\Phi \nabla V^2 + \Delta p=0
\label{de2}
\ee
Let us now suppose the cosmological fluid has a quasi-isotropic and slow-varying pressure. This rough hypothesis could certainly be replaced by some more general extimate on the dependence of $dp$ on $R$. For example the rest of our treatment would be substantially unchanged if we would assume $dp\propto R^{-2}$ and $\Delta p \propto R^{-3}$, but in absence of a concrete physical basis for such extimates, we prefer to simply neglect all terms depending on the derivatives of the pressure. Assuming constant or quasi-constant pressure still does not mean assuming a trivial thermodynamics unless one additionally assumes $dS=0$. Note moreover that in our formal recovering of the Friedmann equations (\ref{Fried1})-(\ref{Fried2}), the thermodynamical variables $\rho_F$ and $p_F$
will be different than our special relativistic analogoues: they will depend on $V^2$, $\nabla (V)^\star$ and $(\nabla V)^\star$ as well as on $\rho$ and $p$. In practice constant $p$ doesn not mean constant $p_F$. This leaves us a certain freedom of choice of hypothesis on the evolution of the special relativistic test-fluid.   
Now from (\ref{V2}) we have: 

\be
\nabla V = 3 \frac{R^\star}{R} -\alpha
\label{divV}
\ee
where we have denoted:

\be
\alpha= \frac{1}{2}\frac{(V^2)^\star}{V^2}
\ee
Replacing $\nabla V$ by (\ref{divV}) in (\ref{X2}) and (\ref{V2}) we have:

\be
\frac{R^\star{}^\star}{R}+\Biggl(\frac{R^\star}{R}\Biggr)^2-\alpha \frac{R^\star}{R} -\frac{V^2}{R^2}
=0
\label{X3}
\ee

\be
\frac{R^\star{}^\star}{R}+2\Biggl(\frac{R^\star}{R}\Biggr)^2-\alpha \frac{R^\star}{R}-\frac{1}{3}(\beta+\alpha^2)
\label{V3}
\ee
where we have denoted:

\be
\beta= (\nabla V)^\star + \nabla (V^\star)
\ee
Taking (\ref{V3}) minus (\ref{X3}) we then have:

\be
\Biggl(\frac{R^\star}{R}\Biggr)^2=\frac{1}{3}(\beta+\alpha^2)
-\frac{V^2}{R^2}
\label{F1}
\ee 
We recognize the same structure of the Friedmann equation (\ref{Fried1}). The corrispondence is only formal, since we have to somehow identify the time derivative with the star derivative, and the scale factor of Friedmann cosmology (which comes from the metric of the curved 4-d spacetime) with our ``typical length". Yet such correspondence is significant. We have that (\ref{F1}) reduces to (\ref{Fried1}) if:

\be
\ba{ll}
\chi \rho_F + \Lambda = \beta +\alpha^2 \cr
k=V^2
\ea
\label{match1}
\ee
In particular our working hypothesis $V_AV^A>0$ leads to $k=1$. It directly gives $k>0$, and exact match with the value $+1$ is not a problem since we still have freedom of choice for the parameter $\mu$: see (\ref{mumu}). 

But it is clear that in general the sign of $k$ is determined by that of $V_AV^A$: we have $k=0$ if $V^2=0$ and $k=-1$ if $V_AV^A=-V^2<0$.  

In particular, from (\ref{mumu}) we see that the case $\epsilon=-1$ (Carmelian relativity) means $V_AV^A<0$. Thus our model in connection with Carmeli's cosmological relativity actually predicts $k=-1$.
 
If instead $\epsilon =+1$ then our model fits with all the three possible values of $k$ and does not select one.

Now replacing $R^\star/R$ from (\ref{F1}) in (\ref{X3}), we have:

\be
\frac{R^\star{}^\star}{R}=-\frac{1}{3}(\beta+\alpha^2)+\frac{V^2}{R^2}+\alpha \sqrt{\frac{1}{3}(\beta+\alpha^2)-\frac{V^2}{2R^2}}
\label{attem}
\ee
By power series expansion in terms of $R^{-1}$ we have:

\be
\sqrt{\frac{1}{3}(\beta+\alpha^2)-\frac{V^2}{2R^2}}= \frac{1}{3}\sqrt{3(\beta+\alpha^2)}-\frac{\sqrt{3(\beta+\alpha^2)}}{4(\beta+\alpha^2)}\frac{V^2}{R^2}+O(\frac{1}{R^4})
\ee
Thus, dropping terms of higher orders, we have from (\ref{attem}):

\be
\frac{R^\star{}^\star}{R}=\frac{1}{3}[\sqrt{3\beta+3\alpha^2}-(\beta+\alpha^2)]+ \Biggl(1-\frac{\sqrt{3(\beta+\alpha^2)}}{4(\beta+\alpha^2)}\Biggr)\frac{V^2}{R^2}
\label{fried2bis}
\ee
This appears to introduce a correction term of order $R^{-2}$ to our analogue to Friedmann equation (\ref{Fried2}).  However we can still match the terms of order zero if:

\be
\frac{\chi}{6}(\rho_F+3p_F)+\frac{\lambda}{3}= \frac{1}{3}[\sqrt{3\beta+3\alpha^2}-(\beta+\alpha^2)]
\label{match2}
\ee   
A solution to system (\ref{match1})-(\ref{match2}) is the following:

\be
\ba{ll}
\chi\rho_F=\beta+\alpha^2 & \chi p_F= (1/3)[\beta+\alpha^2-2\sqrt{\beta+\alpha^2}]
\ea
\ee
The possible additional condition: $\beta=3/16-\alpha^2$ lets the $R^{-2}$ correction vanish, so that both Friedmann equations (\ref{Fried1})-(\ref{Fried2}) are formally recovered, but leads to constant values for $\rho_F$ and $p_F$:  

\be
\ba{ll}
\chi\rho_F=3/16 & \chi p_F= -7/16
\ea
\ee
Thus in this case we have $p_F<0$.

Negative pressure cannot be discarded in cosmology and astrophysics (see e.g.\cite{bonnor60,kunzle67,wesson86,ebert89,katz91}) and even in some hydrodynamical problems (involving turbolence and moving 
boundaries: see e.g. \cite{mana48,greenhow,lif}). Here it is even less surprising, since we are dealing with
cosmological particle production. However, 
our source $-E'$ of particle production is actually independent on the value or sign of $p_F$, i.e. the particle production mechanism considered in section 2 and 3 does not need $p_F<0$ nor $p<0$. 

The formal recovering of the Friedmann equations - with the possible correction displayed by (\ref{fried2bis}) - and 
the condition $k=-1$ obtained by compatibility with Carmeli's cosmological relativity are, in the writer's opinion, very interesting results. However, there is still the problem to understand if such results can be considered compatible with the particle production mechanism considered in section 5 and how, since density and pressure involved in the two cases are different fields.

\section*{Acknowledgements}

I am grateful to Prof. J. G. Hartnett at the 
University of Western Australia
for his help.

\end{document}